\documentclass[]{tPHM2e}

\begin{document}
%\doi{10.1080/1478643YYxxxxxxxx}
%\issn{1478-6443}
%\issnp{1478-6435}
%\jvol{00} \jnum{00} \jyear{2008} \jmonth{00}

\markboth{V. Blavatska, C. von Ferber and Yu. Holovatch}{Scaling of complex polymers}

\articletype{}

\title{Scaling of complex polymers: new universality classes and beyond}

\author{
V. Blavatska $^{\rm a,b}$ $^{\ast}$\thanks{$^\ast$Corresponding author. Email: viktoria@icmp.lviv.ua
\vspace{6pt}}, C. von Ferber$^{\rm c,d}$  and Yu. Holovatch$^{\rm b,e}$ \\
 $^{\rm a}${\em{ Institut f\"ur Theoretische Physik, Universit\"at Leipzig,
D-04103 Leipzig, Germany}}; \\
$^{\rm b}${\em{Institute for Condensed Matter Physics, National
Academy of Sciences of Ukraine, UA--79011 Lviv, Ukraine}}; \\
$^{\rm c}${\em{Applied Mathematics Research Centre, Coventry University,
 Coventry, UK}};\\
 $^{\rm d}${\em{Institute of Physics, Freiburg University,
                D-79104 Freiburg, Germany}};\\
                $^{\rm e}${\em{ Institut f\"ur Theoretische Physik, Johannes Kepler
Universit\"at Linz, A-4040, Linz, Austria}}}
%\vspace{6pt}
%\received{v4.0 released January 2008} }

\maketitle
\begin{abstract}

\bigskip
We analyse scaling laws that govern macromolecules of different
topology: polymer chains, homogeneous and miktoarm star polymers in a
good solvent possibly constrained by a porous medium.  The latter
is modelled by long-range-correlated disorder with a pair correlation
function $g(r)$ that decays with a power law $g(r)\sim r^{-a}$ at
large distances $r$. We show that this type of disorder changes the
universality class of the system. Within the frames of the
field-theoretical renormalization group approach we obtain the
corresponding new universal critical exponents for systems of
homogeneous and star copolymers and discuss different consequences of
the architecture dependent change of the scaling behaviour.
\bigskip

\end{abstract}

\section{Introduction}

It is well established, that the statistics of long flexible polymer
macromolecules in a good solvent is governed by scaling laws in the
asymptotic limit of infinitely long chains. The model of
self-avoiding walks (SAWs) is regarded as most
successful in capturing the universal behaviour of polymer
chains \cite{polymers}. In particular, for the averaged end-to-end
distance and number of configurations of SAW with $N$ steps one
finds:
\begin{eqnarray}
\langle R^2 \rangle {\sim} N^{2\nu_{\rm SAW}},\,\,{\cal Z}_N{\sim} {\rm
e}^{\mu N} N^{\gamma_{\rm SAW}-1}, \,\,\,\,d{=}3:\,\nu_{{\rm
SAW}}{=}0.588(2), \,\gamma_{{\rm SAW}}{=}1.159(6),\label{sawscaling}
\end{eqnarray}
where $\nu_{{\rm SAW}}, \gamma_{{\rm SAW}}$ are universal exponents,
that depend on the space dimension $d$ only (for numerical
estimates see Ref. \cite{Guida98}), $\mu$ is a non-universal
fugacity. The scaling laws (1) correspond to the swelling regime of
of polymer coils. Note, that the situation changes, when
the temperature of the solution decreases. For any generic polymer
solvent system one finds a so-called $\Theta$-temperature \cite{polymers},
at which the solvent mediated attractive and repulsive interactions
 between the monomers
cancel. In this situation the scaling properties of the polymer chain
can be described by that of a
simple random walk (RW). Then, the scaling laws
(\ref{sawscaling}) hold with exponents $\nu_{{\rm RW}}{=}1/2,
\gamma_{{\rm RW}}{=}1$. At temperatures below the $\Theta$-point, the
polymer undergoes a collapse transition and passes to a globule
regime with a scaling exponent $\nu_{{\rm c}}=1/d$.

As established by de Gennes \cite{polymers}, the scaling
properties of SAWs can be studied in the language of field theory,
by mapping the $O(m)$-symmetrical spin model at its
critical point to polymer field theory by a formal $m\to 0$ limit.
In this way, the
powerful renormalization group (RG) methods, which are
traditionally used to quantitatively describe the critical behaviour
of magnetic systems and to determine the critical exponents,
can be applied to polymer solutions. In the frames of the RG, the
scaling behaviour of the model is determined by the stability and
physical accessibility of a corresponding fixed point (FP). For a
polymer chain, the crossover from the RW regime to that of the SAW
corresponds to the interchange of stability between the Gaussian (G) fixed
point (corresponding to absence of interaction between monomers) and
the Polymer (P) fixed point. The scaling of a SAW is governed in
 by exponents (1) that belong to the $O(m=0)$
universality class. These are the most prominent
scaling exponents in polymer physics, however, there are more.
On the one hand, there is a series of
exponents governing the scaling of polymers of complex topology --
star polymers and networks \cite{Likos01,Ferber02,Duplantier89}. On
the other hand, the scaling laws (1) may change if e.g. disorder in the system
gives rise to a separate stable FP. Both cases are considered below.

The present communication aims to attract attention to appearance of
new scaling laws related to a constraint on the
polymer system in the form of a disordered environment or to a more
complex architecture of the polymers.  In particular, we will be
interested in investigating the interplay of different aspects of
complexity appearing simultaneously such as the degree of branching of
the polymer, its possibly inhomogeneous composition and the type of
disorder in the environment.  We derive and calculate the scaling exponents
for the corresponding universality classes and discuss the relevant
observable phenomena.  The outline of this paper is as follows: we
first describe different possible scenarios of scaling behaviour and
present the field-theoretical description for models of polymers in
these scenarios. We then discuss the results and the underlying
physical phenomena.  Our discussion encompasses previous results
\cite{Blavatska01,Blavatska06} as well as recent extensions
\cite{Blavatska09}.

\section{Origin of scaling exponents: composite operators and fixed points}

From the point of view of RG transformation, the scaling exponents
(1) originate from anomalous dimensions of local
operators evaluated at the stable FP.
 The scaling laws can  be altered by
connecting single polymer chains into more complex structures, which
in the language of field theory amounts to the appearance of new
composite operators. The most simple non-trivial representative of
the class of branched polymers is the so-called star polymer, which
consists of linear chains, linked together at one of their end-points. For
a star with $f$ arms each with $N$ steps (monomers), the number of
possible configurations scales as \cite{Schafer-1992}:
\begin{equation}\label{gamstar}
Z_{N,f} \sim {\rm e}^{\mu Nf} N^{\gamma^{}_f-1}\sim(R/\ell )^
{\eta^{}_f-f\eta^{}_2}.
\end{equation}
The second part shows the power law in terms of the size $R$ of a
single chain of $N$ monomers on microscopic step length $\ell$.  The
exponents $\gamma_f$, $\eta_f$ are universal star exponents,
depending on the space dimension $d$ and number of arms $f$.
This includes the single polymer chain with $f=1$ or $2$.
Some results on star polymers  can be found in Refs. \cite{Schafer-1992,Ferber95,Hsu04}.

Linking together polymers of different species, results in
non-homogeneous polymer stars, which may have a richer scaling
behaviour \cite{Schafer91,Ferber97,Ferber99}. A particular case is
the copolymer star, consisting of polymer chains with two different
$\Theta$-temperatures. Let $f_1$ chains be in the good solvent
condition while further $f_2$ chains of the star polymer are at
their  $\Theta$-temperature. Then the partition function of this
star copolymer scales as\cite{Ferber97}:
\begin{equation}\label{comstar}
Z^X_{f_1,f_2} \sim(R/\ell )^
{\eta^X_{f_1,f_2}-f_1\eta^X_{2,0}-f_2\eta^X_{0,2}},
\end{equation}
where $\eta^X_{f_1,f_2}$ represents a family of copolymer star exponents
and $X$ indicates one of eight possible FPs in this ternary polymer scenario
\cite{Schafer91}.
 What is the influence of structural disorder on the scaling laws Eqs.(2),(3)?
Such structural disorder may be present in a solution in terms of a
porous medium. It has been shown however, that uncorrelated
point-like defects of weak concentration will not induce any
change of polymer scaling behaviour \cite{Kim83}: the polymer FP, as for SAW
on a pure lattice, remains stable and physically accessible. More
interesting is  the case of so-called long-range-correlated
disorder, where the distribution of structural defects is
characterised by a correlation function that decays with a power law
$g(r)\sim r^{-a}$ \cite{Weinrib83} for
large $r$. The non-trivial influence of long-range (LR) disorder on
the scaling properties of single polymer chains and star polymers
has recently been worked out in the frames of the field-theoretical approach
\cite{Blavatska01,Blavatska06}.
As mentioned earlier, the statistics of
flexible polymer chains can be extracted in the formal limit $m{\to}0$
of an $m$-vector model at its critical point. Similarly, the
scaling of a polymer experiencing LR correlated disorder
can be studied using the effective Hamiltonian:
\begin{eqnarray}
\lefteqn{
{\cal H}_{eff}{=}\sum_{k}  \left[ \frac{1}{2}\sum_{\alpha=1}^n
(k^2 + \mu_0^2)\,\vec {\varphi}^{\alpha}_k\vec
{\varphi}^{\alpha}_{-k}\right]\label{lag}
{+}\sum_{k_1,k_2,k_3,k_4}\delta(k_1+k_2+k_3+k_4)\times
}\\
&&\times\left[\sum_{\alpha=1}^n
\frac{u_0}{4!}
\vec{\varphi}^{\alpha}_{k_1}\vec{\varphi}^{\alpha}_{k_2}
\vec{\varphi}^{\alpha}_{k_3}\vec{\varphi}^{\alpha}_{k_4}\right.
%\nonumber\\&&
{-}\left.\frac{w_0}{4!}\sum_{\alpha,\beta=1}^n |k_1-k_2|^{a-d}
\vec{\varphi}^{\alpha}_{k_1}\vec{\varphi}^{\alpha}_{k_2}
\vec{\varphi}^{\beta}_{k_3}\vec{\varphi}^{\beta}_{k_4}\right].\nonumber
\end{eqnarray}
Here, $\vec {\varphi}^{\alpha}_k$ is an $m$-component vector field,
$u_0$, $w_0,$, $\mu_0$ are bare couplings and mass, Greek indices
denote replicas and both the replica limit $n\to 0$ and $m\to 0$ are implied.

The impact of disorder on the universal behaviour can be singled out by
analysing the occurrence and stability of the disorder induced
fixed point. The appearance of a new universality class within the model
(\ref{lag}) was discussed in Ref. \cite{Blavatska01} using two
complementary RG approaches: a first order expansion in `small' parameters
$\varepsilon=4-d$ and $\delta=4-a$ following Ref.\cite{Weinrib83} and a
second order calculation using numerically evaluated integrals at fixed
$d$ and $a$ of Ref.\cite{Prudnikov1999}. Our latter two loop calculation
confirms the existence and stability of a LR disorder induced FP for polymers.
However, the FP as found by the
first order $\varepsilon, \delta$-expansion appeared to be
stable in a non-physical region.

Here, it is our intention to investigate a higher order approximation
to confirm the stability of the LR disorder FP in the physical region
also within the $\varepsilon,\delta$-expansion.
We use the second-order RG $\beta$-functions of the $m$-component model in
LR correlated disorder, as derived in Ref. \cite{Korutcheva98}.
Converting these to continuous differential equations, and passing to the $m\to
0$ limit we find:
\begin{eqnarray}
\frac{{\rm d} u}{{\rm d} \kappa}&=&-2\eta\,u+\epsilon
u+6\,uw-4{u}^{2}- 2{w}^{2}-30{u}^{2}w+ 27\,{w}^{2
}u-8\,{w}^{3}+11{u}^{3},\nonumber\\
\frac{{\rm d} w}{{\rm d} \kappa}&=&-2\,\eta\,w+
\delta\,w-2\,uw+2\,{w}^{2}+3/2\,{u}^{2}w+3\,{w}^{3}-6\,{w}^{2}u,
\end{eqnarray}
where $\kappa$ is a rescaling parameter,
$\eta=\delta(2\varepsilon-\delta)/16$.
Looking for the FPs of this RG flow, we find the LR disorder induced FP $
u^{LR}={\delta^2(8\,\varepsilon^3-24\,\delta\,\varepsilon^2+24\,\delta^2\,\varepsilon-8\,\delta^3+14\,\delta\,\varepsilon^3-18\,\delta^2\,\varepsilon^2+4\,\delta^3\,\varepsilon+\delta^4-4\,\varepsilon^4)}/{16\,(\varepsilon-\delta)^4}$,
$w^{LR}=\delta(40\,\delta\,\varepsilon^3-21\,\delta\,\varepsilon^4-72\,\delta^2\,\varepsilon^2+56\,\delta^3\,\varepsilon-16\,\delta^4-5\,\delta^5-77\,\delta^3\,\varepsilon^2+36\,\delta^4\,\varepsilon-8\,\varepsilon^4$+$2\,\varepsilon^5+62\,\delta^2\,\varepsilon^3)/16\,(\varepsilon-\delta)^4$.
This FP is
 stable and physically accessible for $\delta>2\varepsilon$ or equivalently $a<2d-4$.
Taking into account, that according to Eq. (\ref{lag}) LR disorder should be relevant for $a<d$,
these second order results establish the presence of the LR fixed point in
the physical region of ($d,a$).

As explained in Ref.\cite{Schafer-1992},
the study of $f$-arm star configurations in this theory is given in
terms of the anomalous dimensions of complex composite operators,
containing products of $f$ zero-component fields.

\section{Results and perspectives}
\begin{figure}
\begin{center}
%\begin{minipage}{100mm}
%\subfigure[]
{
\hspace*{1.6cm}
\resizebox*{8cm}{!}{\includegraphics{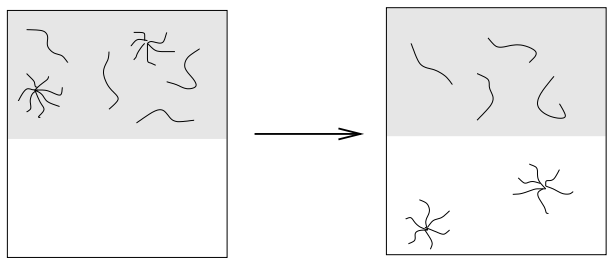}}}
\caption{Separation of linear and star polymers in solution, part of which is in a porous medium (shown in grey).}
\label{star}
%\end{minipage}
\end{center}
\end{figure}

The presence of a new stable fixed point leads to different
values of the scaling exponents for polymers. In particular, for star polymers, the first order expressions
for the exponents $\eta_f,\gamma_f$ for a star in a pure solvent and
for a star influenced by long-range correlated disorder (LR) read:
\begin{eqnarray}
&&\eta_f^{(LR)}= -\frac{1}{4}\delta f(f-1),\,\,\,\,\,\,\,\eta_f^{(pure)}=
-\frac{1}{8}\varepsilon f(f-1),\\
&& \gamma_f^{(LR)}= 1-\frac{1}{8}\delta f (f-3), \,\,\, \gamma_f^{(pure)}=
1-\frac{1}{16}\varepsilon f(f-3).\label{change}
\end{eqnarray}
 Note the decrease of $\gamma_f$ at fixed $f>3$ with increasing $\delta$ (i.e. when the
 correlation of the disorder becomes stronger). However, the behaviour for chain
polymers $f=1,2$ differs: in this case the exponents
$\gamma_1=\gamma_2$ increase for increasing $\delta$.
Omitting factors of $kT$, let ${\cal F}^{pure}_f(N)=\ln {\cal Z}^{pure}_f(N)$ be the free energy of a star polymer in the pure solvent
 and ${\cal F}^{(LR)}=\ln {\cal Z}^{LR}_f(N)$ its free energy
in a porous medium. Then we can estimate the free energy difference of a star polymer with respect to the two environments:
\begin{eqnarray}\nonumber
{\cal F}^{LR}_f(N)-{\cal F}^{pure}_f(N)&=& (\mu^{LR}-\mu^{pure})fN+(\gamma_f^{LR}-\gamma_f^{pure} )\ln N.
\label{osmos2}
\end{eqnarray}
The difference in the free energy consists of two terms. The first is
independent of the polymer architecture and simply takes into account
the change in chemical potential per monomer. Two polymers of the same
molecular weight (same total number of monomers $fN$) will experience the same
difference. The second term which grows with $\ln N$ is architecture
dependent. Following Eqs. (\ref{change}),  we see that this term has different
signs for chain polymers ($f=1, 2$), and star polymers ($f\geq 3$).
Thus, even though both
may be expected to be expelled from the correlated environment, this effect
is weakened for linear polymers and enhanced for star polymers of the same
molecular weight. The first term could in principle be cancelled out if the
pure part of the solution is exchanged by a second porous medium which however
displays short range (SR) disorder with the same chemical potential
$\mu^{SR}=\mu^{LR}$. Static architecture induced separation results, see Fig.\ref{star}.
%---
%---

Exploring further aspects of complexity, we consider star polymers
built from chains of species with different $\Theta$-temperatures
(copolymer stars).
Depending on the temperature, one or more of inter- and intra-chain
interactions may vanish, and additional FPs appear for the inter-chain
interaction. A species with vanishing intra-chain interaction is
represented by RWs avoiding the other species.  This situation may
also be used to describe the diffusion of particles near an absorbing
polymer \cite{Cates87}.  The moments of the resulting
flow near the polymer turn out to be multi-fractal \cite{Cates87,Ferber99}.
The scaling exponents of copolymer stars also determine the scaling of
corresponding copolymer networks \cite{Duplantier89,Schafer-1992,Ferber97}
and play role in a number of other phenomena, given below.

Treating the model of star copolymers in a porous medium by the RG
approach, we have found different scenarios of scaling behaviour and
corresponding values for scaling exponents \cite{Blavatska09}. To
summarise some results for the exponents, let us first introduce some
notations. Each regime is denoted by two letters, corresponding to
species 1 and 2 respectively and reflecting, whether it behaves
like random walk ($R$) or self-avoiding walk $S$. The index
$(\cdots)_L$ denotes the presence of long-range-correlated disorder.
The index $(\cdots)_i$ corresponds to the situation, when species
$1$ and $2$ interact with each other. When disorder is absent, we
reproduce known results \cite{Schafer91} for the cases of two
interacting random walks $(RR)_i$, two interacting SAWs $(SS)_i$ and
SAWs interacting with random walks $(SR)_i$:
\begin{eqnarray}
\eta_{f_1,f_2}^{(RR)_i}&{=}&\frac{-(f_1f_2)\varepsilon}{2},\\
\eta_{f_1,f_2}^{(SS)_i}&{=}&\frac{-(f_1+f_2)(f_1+f_2-1)\varepsilon}{8},\\
\eta_{f_1,f_2}^{(SR)_i}&{=}&\frac{-f_1(f_1+3f_2-1)\varepsilon}{8}.
\end{eqnarray}
 $\eta_{f_1,f_2}^{(SS)_i}$ is equivalent to the exponent of a homogeneous star polymer of $f=f_1+f_2$ arms.
Turning on the long-range-correlated disorder, we obtain estimates for the exponent
$\eta_{f_1,f_2}$ for different regimes in a new universality class:
\begin{eqnarray}
\eta^{(RR)_{iL}}_{f_1,f_2}&{=}&-(f_1f_2)\delta,\\
\eta^{(SS)_{iL}}_{f_1,f_2}&{=}& \frac{-(f_1+f_2)(f_1+f_2-1)\delta}{4},\\
\eta^{(SR)_{iL}}_{f_1,f_2}&{=}&\frac{-f_1(f_1+3f_2-1)\delta}{4}.
\end{eqnarray}
Here, $\eta^{(S_LS_L)_{iL}}_{f_1,f_2}$ is the exponent for the homogeneous star
with $f_1+f_2$ arms in long-range-correlated disorder,
 $\eta^{(RR)_{iL}}_{f_1,f_2}$ and $\eta^{(SR)_{iL}}_{f_1,f_2}$ describe $f_2$ random walks,
interacting with $f_1$ RWs and SAWs respectively,
in LR disorder.

Besides the usual polymer exponents  describing statistical
properties like end-to-end distance or number of configurations,
polymers of complex structure (e.g. star copolymers) are
characterised by the full set of star exponents. These star
exponents form a basis to determine the scaling of complex copolymer
networks \cite{Duplantier89,Schafer-1992,Ferber97}. They are
important for a wide range of phenomena, some examples are given by
the absorption of diffusive particles  in the vicinity of an
absorbing polymer \cite{Cates87,Ferber99}, the effective interaction
in colloids of star polymers \cite{Likos01}, the transition of
double stranded to single stranded DNA  \cite{Baiesi02}. As we have
demonstrated in this paper polymer scaling behaviour is influenced
in a non trivial way by correlated structural disorder. Further
investigations are under way to complete the quantitative
description of the scenarios of complex polymer behaviour in
correlated environments.
\section*{Acknowledgement}
We acknowledge support by  a ``Marie Curie International Incoming
Fellowship" (V.B.) and Austrian FWF under Project No. P19583-N20
(Yu.H).

\medskip
\markboth{V. Blavatska, C. von Ferber and Yu. Holovatch}{Scaling of complex polymers}


\begin{thebibliography}{20}

\markboth{V. Blavatska, C. von Ferber and Yu. Holovatch}{Scaling of complex polymers}

 \bibitem[1]{polymers}
See, e.g.: G.-P. de Gennes, {\em Scaling concepts in polymer Physics},
Cornell University Press, Ithaca and London, 1979; L.  Sch\"afer, {\em  Universal Properties of Polymer
Solutions as Explained by the Renormalization Group},
Springer, Berlin, 1999.


\bibitem[2]{Guida98}
R. Guida, J. Zinn Justin, {\em Critical exponents of the $N$-vector model},  J. Phys. A 31 (1998), pp.~8103-8121.


\bibitem[3]{Likos01}
C.N. Likos, {\em Effective interactions\! in soft \! condensed matter physics},  Phys. Rep. 348 (2001), pp.~267-439.

\bibitem[4]{Ferber02}
C. von Ferber and Yu. Holovatch (eds.) {\it Special Issue ``Star Polymers"}, vol. 5, Condens. Matter Physics (2002).



\bibitem[5]{Duplantier89}
B. Duplantier,   {\em   Statistical mechanics of polymer networks of any topology}, J. Stat. Phys.  54 (1989),
  pp. 581-680.

\bibitem[6]{Blavatska09}
V. Blavatska, C. von Ferber, Yu. Holovatch, to be published.


 \bibitem[7]{Blavatska01}
V. Blavatska,  C. von Ferber and  Yu. Holovatch, {\em Polymers in
media with long-range-correlated quenched disorder},  J. Mol. Liq.
91 (2001), pp. 77-84; {\em Polymers in long-range-correlated
 disorder},  Phys. Rev. E 64 (2001), pp. 041102(1-10);

 \bibitem[8]{Blavatska06}
V.  Blavatska,  C. von Ferber, and  Yu. Holovatch {\em
Entropy-induced separation of star polymers in porous media}, Phys.
Rev. E  74 (2006), pp. 031801(1-12).

\bibitem[9]{Schafer-1992}
L. Sch\"afer et al., {\em  Renormalization of polymer networks and stars},  Nucl. Phys. B  374 (1992), pp.~473-495.



 \bibitem[10]{Ferber95}
 C. von Ferber and  Yu. Holovatch, {\em Star exponents in polymer theory: renormalization group results in three dimensions},  Condens. Matter Phys. 5 (1995), pp. 8-22; {\em Polymer stars in three dimensions. Three-loop results},  Theor. Math. Physics 109 (1996), pp. 1274-1286.


 \bibitem[11]{Hsu04}
H.P.  Hsu,  W. Nadler  and  P. Grassberger, {\em Scaling of Star Polymers with 1-80 Arms},   Macromolecules~ 37 (2004),  pp. 4658-4663.

 \bibitem[12]{Schafer91}
 L. Schaefer,  U. Lehr and  C. Kapeller, {\em Higer order calculations of the renormalization group flow for multicomponent polymer solutions}, J. Phys. I 1 (1991), pp. 211-233.

 \bibitem[13]{Ferber97}
 C. von Ferber and  Yu. Holovatch  {\em Copolymer networks and stars: Scaling exponents},  Phys. Rev. E  56 (1997), pp. 6370-6386; {\em Copolymer networks: Multifractal dimension spectra in polymer field theory},  Europhys. Lett. 39 (1997), pp. 31-36.




 \bibitem[14]{Ferber99}
  C. von Ferber   and  Yu. Holovatch,  {\em Multifractality of
  Brownian motion near absorbing polymers},   Phys. Rev. E  59 (1999), pp. 6914-6923.

\bibitem[15]{Kim83}
Y. Kim,
{\em Renormalization-group study of self-avoiding walks on the random lattice}, J. Phys. C 16 (1983), pp. 1345-1352.

\bibitem[16]{Weinrib83}
 A. Weinrib  and  B.I. Halperin,   {\em Critical phenomena in systems with long-range-correlated quenched disorder},
   Phys. Rev. B 27 (1983), pp.  413-427.

\bibitem[17]{Prudnikov1999}
V. V. Prudnikov, P. V. Prudnikov, and A. A. Fedorenko,
J. Phys. A-Math. Gen. 32 (1999), pp. 8587-8600; V. V. Prudnikov and A. A. Fedorenko ibid. pp. L399-L405

\bibitem[18]{Korutcheva98}
E. Korutcheva, F. de la Rubia, {\em Dynamival properties of the Landau-Ginzburg model with long-range correlated quenched impurities},
Phys. Rev. B {\bf 58} (1998), pp. 5153-5156.

  \bibitem[19]{Cates87}
  M.E. Cates   and  T.A. Witten,   {\em Diffusion near absorbing fractals:
  Harmonic measure exponents for polymers}, Phys. Rev. A 35 (1987), pp. 1809-1824.

\bibitem[20]{Baiesi02}
M. Baiesi, E. Carlon, A.L. Stella, {\em Scaling in DNA unzipping
models: Denaturated loops and end segments as branches of a block
copolymer network}, Phys. Rev. E {\bf 66} (2002), pp. 021804 (1-8).

\end{thebibliography}
\end{document}